\documentclass[aps,pre,twocolumn,showpacs,floatfix]{revtex4} 

\usepackage{amsmath,bm,epsfig}

\begin{document}

\title{Refined similarity hypotheses in shell models of turbulence}
\author{Emily S. C. Ching$^{1,2}$, H. Guo$^1$, and T.S. Lo$^1$}
\affiliation{$^1$ Department of Physics,
The Chinese University of Hong Kong, Shatin, Hong Kong}
\affiliation{$^2$ 
Institute of Theoretical Physics,
The Chinese University of Hong Kong, Shatin, Hong Kong}

\date{\today}
 
\begin{abstract}
A major challenge in turbulence research is to understand from first principles
the origin of anomalous scaling of the velocity fluctuations in high-Reynolds-number
turbulent flows. 
One important idea was proposed by Kolmogorov [J. Fluid Mech. {\bf 13}, 82 (1962)],
which attributes the anomaly to the variations of the locally averaged energy dissipation rate.
Kraichnan later pointed out [J. Fluid Mech. {\bf 62}, 305 (1973)] that the locally averaged
energy dissipation rate is not an inertial-range quantity and a proper inertial-range quantity 
would be the local energy transfer rate. As a result, Kraichnan's idea attributes 
the anomaly to the variations of the local energy transfer rate. These ideas, generally known 
as refined similarity hypotheses, can also be extended to study the anomalous scaling of 
fluctuations of an active scalar, like the temperature in turbulent convection. In this paper, 
we examine the validity of these refined similarity hypotheses and their extensions to an active 
scalar in shell models of turbulence. 
We find that Kraichnan's refined similarity hypothesis and its extension are valid.
\end{abstract}
\maketitle

\section{Introduction}

Much effort in turbulence research has been devoted to the study 
of the possible universal statistics of velocity fluctuations 
in turbulent flows at high Reynolds number. The seminal work of Kolmogorov
in 1941 (K41)\cite{K41} introduced the idea of universal homogeneous and isotropic statistics
in high-Reynolds-number turbulent flows and in particular deduced that the statistical 
moments of the velocity differences, 
$\delta u_r(\vec{x},t) \equiv |\vec{u}(\vec{x}+\vec{r},t)-\vec{u}(\vec{x},t)|$, 
where $\vec{u}$ is the velocity field, have power-law scaling with the 
separating distance $r = |\vec{r}|$:
\begin{equation}
\langle \delta u_r(\vec{x},t)^n \rangle \sim
(\langle \epsilon \rangle r)^{n/3}
\label{K41}
\end{equation}
when $r$ is within the inertial range. 
Here $\langle \ldots \rangle$ denotes an ensemble average. The inertial range refers to 
the range of length scales that are smaller than those of energy input and larger 
than those affected directly by molecular dissipation, and the energy dissipation 
rate per unit mass $\epsilon$ is given by 
\begin{equation}
\epsilon(\vec{x},t) = 
\frac{\nu}{2} \sum_{i=1}^{3} \sum_{j=1}^{3} \left[\frac{\partial u_i(\vec{x},t)}{\partial x_j} + 
\frac{\partial u_j(\vec{x},t)}{\partial x_i}\right]^2
\label{epsilon}
\end{equation}
where $\nu$ is the kinematic viscosity, $\vec{x} = (x_1,x_2,x_3)$, and
$\vec{u}=(u_1,u_2,u_3)$.
The direct proportionality of the scaling exponents $n/3$ with the order $n$ 
of the statistical moments is equivalent to an independence of $r$ of
the functional form of the probability density function of $\delta u_r$.
Experimental observations confirm the power-law scaling but indicate that the 
scaling exponents depend on the order $n$ in a nonlinear fashion. 
This deviation of the velocity scaling behavior from the K41 results is 
known as anomalous scaling and implies that turbulent velocity fluctuations
have scale-dependent statistics and are thus intermittent.

A major challenge is to understand, from first principles, the origin 
of anomalous scaling. In his refined theory in 1962~\cite{RSH},
Kolmogorov replaced the global average energy 
disipation rate $\langle \epsilon \rangle$
by the locally averaged energy dissipation rate $\epsilon_r$, given by
\begin{equation}
\epsilon_r(\vec{x},t) = \frac{3}{4\pi r^3} \int_{|\vec{h}|\le r} \epsilon(\vec{x}+\vec{h},t) d \vec{h}
\label{epsilonr}
\end{equation}
The possible dependence of $\langle \epsilon_r^{n/3} \rangle$ on $r$ allows for a correction to the
$n/3$ scaling~(see Sec.~\ref{review}). 
As a result, this idea of Kolmogorov, which we refer to as Kolmogorov's refined
similarity hypothesis (RSH) attributes
the origin of anomalous scaling to the
variations of the local energy dissipation rate. 
Kraichnan~\cite{Kraichnan} later pointed out that the local 
energy dissipation rate $\epsilon_r$ is not an inertial-range quantity and a proper
inertial-range quantity would be the local energy transfer rate,
$\Pi_r(\vec{x},t)$, 
which measures the rate of energy transfer into scales 
smaller than $r$ at a local point $\vec{x}$ in space. 
In other words,
Kraichnan's idea, which we refer to as Kraichnan's RSH, 
attributes the origin of anomalous scaling to the
variations of the local energy transfer rate.
In statistically steady state, for $r$ in the inertial range, 
$\langle \Pi_r \rangle = \langle \epsilon \rangle = \langle \epsilon_r \rangle$
but $\Pi_r(\vec{x},t) \ne \epsilon_r(\vec{x},t)$. 
Hence the two RSHs of Kolmogorov and Kraichnan would
give different results for the anomalous scaling exponents of velocity structure functions.
 
Anomalous scaling behavior has also been observed in the statistics of a scalar 
field advected by a turbulent velocity field.
A passive scalar leaves the velocity statistics intact while an
active scalar couples with the velocity
and influences its statistics. The nonlinear problem of
anomalous scaling of active scalars, like that of velocity, remains unsolved.
A common example of an active scalar is temperature in turbulent convection
in which temperature variations drive the flow.
Both Kolmogorov and Kraichnan's RSHs can be extended to turbulent convection
to account for the anomalous scaling behavior of an active scalar.

In this paper, we examine the validity of Kolmogorov and Kraichnan's RSHs
and their extensions to turbulent convection in shell models of turbulence.
Shell models for homogeneous and isotropic turbulence 
have been proved to be very successful in reproducing many
of the statistical features of turbulent flows, and in which high Reynolds number
can be achieved with relative ease~\cite{Shell}.
Shell models for homogeneous turbulent convection have also been studied.
We use two shell models: the Sabra model for homogeneous and isotropic
turbulence~\cite{Sabra}, which is an improved version of the so-called 
Gledzer-Ohkitani-Yamada (GOY) model~\cite{G73,YO87}, and the Brandenburg model
for homogeneous turbulent convection~\cite{Brandenburg}.

The rest of this paper is organized as follow. We first summarize 
the mathematical formulations of the RSHs of Kolmogorov and Kraichnan and the previous
studies that examined their validity in Sec.~\ref{review}. 
In Sec.~\ref{extension}, we discuss how 
these RSHs are extended to turbulent convection in which 
temperature is an active scalar. We describe the Sabra shell model for homogeneous 
and isotropic turbulence and the Brandenburg shell model for homogeneous turbulent convection
in Sec.~\ref{Sabra} and Sec.~\ref{Brandenburg}, respectively.
In Sec.~{\ref{result}}, we examine the validity of the RSHs of Kolmogorov and Kraichnan for
homogeneous and isotropic turbulence in the Sabra model and the validity of their extensions 
to turbulent convection in the Brandenburg model.
Finally, we conclude in Sec.~{\ref{conclusion}}.

\section{Kolmogorov and Kraichnan's 
refined similarity hypotheses}
\label{review}

Kolmogorov's RSH can be stated as:
\begin{equation}
\delta u_r(\vec{x},t) = \hat{\phi} \ [\epsilon_r(\vec{x},t) r]^{1/3}
\label{KolRSH}
\end{equation}
where $\hat \phi$ is a random variable independent of $r$ and statistically independent of 
$\epsilon_r$ when Reynolds number is much larger than 1.
Equation~(\ref{KolRSH}) implies that
\begin{equation}
\langle (\delta u_r)^n \rangle \sim \langle \epsilon_r^{n/3} \rangle r^{n/3}
\label{structureKol}
\end{equation}
For homogeneous flows, $\langle \epsilon_r(\vec{x},t) \rangle = \langle \epsilon(\vec{x},t) \rangle$
and is independent of $r$ and $\vec{x}$.
For $n \ne 3$, $\langle \epsilon_r^{n/3} \rangle$ generally depends on
$r$ and this $r$-dependence allows for a correction to the K41 scaling. Thus
Kolmogorov's RSH attributes the origin of anomalous scaling to the
variations of the local energy dissipation rate.

Similarly, Kraichnan's RSH can be stated as
\begin{equation}
\delta u_r(\vec{x},t) = \hat{\psi} \ [|\Pi_r(\vec{x},t)| r]^{1/3}
\label{KraRSH}
\end{equation}
where $\hat \psi$ is a random variable independent of $r$ and
statistically independent of $\Pi_r$. The local energy transfer rate 
$\Pi_r$ can be defined using banded Fourier series~\cite{Kraichnan}
Equation (\ref{KraRSH}) then implies
\begin{equation}
\langle (\delta u_r)^n \rangle \sim \langle |\Pi_r |^{n/3} \rangle r^{n/3}
\label{structureKra}
\end{equation}
and Kraichnan's RSH thus attributes the origin of anomalous scaling to
the variations of the local energy transfer rate.
As the scaling behavior of $\langle |\Pi_r|^{n/3} \rangle$ would be generally different
from that of $\langle \epsilon_r^{n/3} \rangle$, the two RSHs,~Eqs.(\ref{KolRSH}) and (\ref{KraRSH}),
would give different results for the anomalous scaling exponents of the velocity structure functions.

Kolmgorov's RSH has been widely used in the
discussions of anomalous scaling
of the velocity structure functions.
There have been quite a number of experimental and numerical studies
that examine the validity of Eq.~(\ref{KolRSH}). Most of these
studies checked whether
$\delta u_r$ and $\epsilon_r$ are correlated.
Statistical correlation between $\delta u_r$ and $\epsilon_r$
or its one-dimensional surrogate which represents $\epsilon$ by
$\nu (\partial u_x/\partial x)^2$ or $(\nu/2) (\partial u_y/\partial x)^2$ was
indeed reported~\cite{SKS92,P92,TV92,CDKS93,PPH97}.
It was, however, noted~\cite{SS94,CDKW95} that at least part of such
a statistical correlation results from kinematical constraints
independent of Navier-Stokes dynamics.
Moreover, it would be more direct to check
the implication of Eq.~(\ref{KolRSH})
that the conditional velocity structure functions at fixed values of
the $\epsilon_r$ would be given by:
\begin{equation}
\langle (\delta u_r)^{n} \ \big| \  \epsilon_r \rangle \sim r^{n/3}
\epsilon_r^{n/3}
\label{condVKol}
\end{equation}
Calculations of such conditional velocity structure functions (using the longitudinal velocity
difference instead of the whole velocity difference) were carried out using direct
numerical simulations of isotropic turbulence~\cite{WCBW96} but
these simulations were limited to moderate Reynolds
numbers making it difficult to draw definitive conclusions.
In particular, a clear demonstration of a
scaling behavior of $r^{n/3}$ at fixed $\epsilon_r$ or
a dependence of $\epsilon_r^{n/3}$ at fixed $r$
is lacking.

On the other hand, the validity of Kraichnan's RSH is much less examined.
In a high resolution direct numerical simulation of isotropic turbulence~\cite{CCEH03},
it was shown that the scaling exponents
of $\langle |r \Pi_r|^{p/3} \rangle$ are close to those of the $p$-order
transverse velocity structure functions. Since the
transverse velocity difference is more intermittent than the
longitudinal velocity difference~\cite{CSNC97}, this result implies that
the scaling exponents of $\langle |r \Pi_r|^{p/3} \rangle$ are close to those
of the $p$-order whole velocity structure functions and is
thus consistent with Eq.~(\ref{structureKra}).
However, a direct examination of Eq.~(\ref{KraRSH}) by studying the
behavior of the conditional velocity structure functions at fixed values of
the local energy transfer rate is yet to be performed.

\section{Refined similarity hypotheses for turbulent convection}
\label{extension}

In turbulent convection, 
the statistics of $\delta u_r$ as well as those 
of the temperature difference $\delta T_r \equiv T(\vec{x}+\vec{r},t)-T(\vec{x},t)$, 
where $T(\vec{x},t)$ is the temperature field, are of interest. 
Buoyancy is expected to drive the dynamics, and thus affecting
the statistics of $\delta u_r$ and $\delta T_r$, for length scales $r$ greater than
the Bolgiano scale~\cite{B59,O59}. 
When buoyancy is dominant, it was suggested~\cite{Lvov} that the statistics are governed by a cascade
of temperature variance, which is proportional to entropy in Boussinesq approximation~\cite{Landau}.
Thus the role of the energy dissipation rate $\epsilon$ is now played by 
the temperature (variance) dissipation rate or the entropy dissipation rate $\chi$, given by
\begin{equation}
\chi = \kappa \sum_{i=1}^{3} \left( \frac{\partial T}{\partial x_i} \right)^2
\end{equation}
where $\kappa$ is the thermal diffusivity of the fluid.
In the same spirit of deriving Eq.~(\ref{KolRSH}), 
$\delta u_r$ and $\delta T_r$ are expressed as 
functions of $\alpha g$, $\chi_r$ and $r$ only, where
$\alpha$ is the volume expansion coefficient of the fluid, $g$ is the acceleration due to
gravity, and $\chi_r$ is the locally averaged entropy dissipation rate, given by
\begin{equation}
\chi_r(\vec{x},t) = \frac{3}{4\pi r^3} \int_{|\vec{h}|\le r} \chi(\vec{x}+\vec{h},t) d \vec{h}
\end{equation}
As usual, the functional dependence is obtained by dimensional analysis and the results are:
\begin{eqnarray}
\delta u_r &=& {\hat \Phi_u} (\alpha g)^{2/5} [\chi_r(\vec{x},t)]^{1/5} r^{3/5}
\label{KolRSHActiveU}\\
\delta T_r &=& {\hat \Phi_T} (\alpha g)^{-1/5} [\chi_r(\vec{x},t)]^{2/5} r^{1/5}
\label{KolRSHActiveT}
\end{eqnarray}
Here $\hat \Phi_u$ and $\hat \Phi_T$ are random variables independent of $r$ and 
statistically independent of
$\chi_r$. Equations~(\ref{KolRSHActiveU}) and (\ref{KolRSHActiveT}) are the 
extensions~\cite{Ching2007} of
Kolmogorov's RSH~[Eq.~(\ref{KolRSH})] to turbulent convection where buoyancy is
dominant, and they attribute the origin of the
anomalous scaling to the variations of the local entropy dissipation rate~\cite{ChingChau}.
In particular, the conditional velocity and temperature structure functions at fixed values of
$\chi_r$ would have simple scaling behavior in $r$:
\begin{eqnarray}
\langle (\delta u_r)^{n} \ \big| \  \chi_r \rangle &\sim& r^{3n/5} 
\label{condKolU}\\
\langle |\delta T_r|^{n} \ \big| \  \chi_r \rangle &\sim& r^{n/5} 
\label{condKolT}
\end{eqnarray} 
given by that of Bolgiano-Obukhov~(BO)~\cite{B59,O59}.

To get the corresponding expressions for the extensions of Kraichnan's RSH~[Eq.~(\ref{KraRSH})]
to turbulent convection, one replaces $\chi_r$ by the locally entropy 
transfer rate $\Pi^{\theta}_r$, 
which is defined as the rate of entropy transfer into scales smaller than $r$ at a local 
point $\vec{x}$ in space. The results thus read:
\begin{eqnarray}
\delta u_r &=& {\hat \Psi_u} (\alpha g)^{2/5} |\Pi^{\theta}_r(\vec{x},t)|^{1/5} r^{3/5}
\label{KraRSHActiveU}\\
\delta T_r &=& {\hat \Psi_T} (\alpha g)^{-1/5} |\Pi^{\theta}_r(\vec{x},t)|^{2/5} r^{1/5}
\label{KraRSHActiveT}
\end{eqnarray}
Here $\hat \Psi_u$ and $\hat \Psi_T$ are random variables independent of $r$ and statistically independent of
$\Pi^{\theta}_r$. Equations~(\ref{KraRSHActiveU}) and (\ref{KraRSHActiveT}) therefore attribute
the origin of anomalous scaling to variations of the local entropy transfer rate.
Their implications are that the conditional velocity and temperature structure functions at fixed values of
$\Pi^{\theta}_r$ would have simple BO scaling:
\begin{eqnarray}
\langle (\delta u_r)^{n} \ \big| \  \Pi^{\theta}_r \rangle &\sim& r^{3n/5}
\label{condKraU}\\
\langle |\delta T_r|^{n} \ \big| \  \Pi^{\theta}_r \rangle &\sim& r^{n/5}
\label{condKraT}
\end{eqnarray}

Turbulent convection is often investigated experimentally
in Rayleigh-B{\'e}nard convection cells heated from below and cooled on
top~(see, e.g.,~\cite{Siggia,GL,Kadanoff} for a review). Such confined
convective flows are highly inhomogeneous in which thermal and viscous boundary layers
are present near the top and the bottom of the cell, and coherent structures
are present. It was argued~\cite{Ching2007} that the presence of 
buoyant flow structures known as plumes could affect the scaling behavior,
causing BO scaling to be invalid. Indeed there were indications from direct 
numerical simulations~\cite{Verzicco} 
and analyses of experimental data~\cite{JoT}
that the scaling behavior of the central region
of confined turbulent convection is not well 
described by BO scaling plus corrections.
We note that in Ref.~\cite{JoT},
the statistics of $\delta u_\tau \equiv |\vec{u}(\vec{x},t+\tau)-\vec{u}(\vec{x},t)|$ 
and $\delta T_\tau \equiv T(\vec{x},t+\tau)-T(\vec{x},t)$, the temporal counterparts 
of $\delta u_r$ and $\delta T_r$, were studied because only measurements taken as a 
function of time at a fixed point in space were available. The validity 
of Eq.~(\ref{KolRSHActiveT})
was explored~\cite{ChingChau} by studying the conditional statistics of 
$\delta T_\tau$ at fixed values of $\chi_r$, estimated by 
$\chi_\tau \propto (1/\tau) \int_{t}^{t+\tau} \kappa (\partial T/\partial t')^2 dt'$.
It was found that for scales larger than the Bolgiano scale,
$\langle |\delta T_\tau|^n | \chi_\tau \rangle/(\langle |T_\tau|^2 | \chi_\tau
\rangle)^{n/2}$ become independent of $\tau$. This result indicates that
the scale- or $\tau$-dependence of the statistics of $\delta T_\tau$ 
can be attributed to the variations of $\chi_\tau$, and that
the extension of Kolmogorov's RSH holds in 
the buoyancy-driven regime in turbulent Rayleigh-B{\'e}nard convection.

On the other hand, 
the Brandenburg shell model for homogeneous turbulent convection is, 
by construction, free of boundaries and thus plumes.
In an earlier work~\cite{ChingCheng},
we have studied the anomalous scaling of an active scalar in the
Brandenburg model and found that the scaling behavior 
is BO plus corrections due to the shell-to-shell variations of the 
entropy transfer rate. These results thus verify the validity 
of the extension of Kraichnan's RSH. 
We shall discuss these results in greater depth and also examine 
the validity of the extension of Kolmogorov's RSH 
in Sec.~\ref{Brandenburg}.

\section{Shell model for homogeneous and isotropic turbulence}
\label{Sabra} 
The basic idea of a shell model is to consider the velocity variable 
$u_n$, which can be associated with the velocity difference $\delta u_r$ with $r=1/k_n$, 
in discrete ``shells"  in Fourier space.
Here $k_n=k_0 \lambda^{n}$, with $n=0,1, \ldots, N-1$, is the 
wavenumber of the $n$th shell, and $\lambda$ is customarily 
taken to be 2.
In the Sabra model~\cite{Sabra}, $u_n$ is complex and satisfies the following
equation of motion:
\begin{eqnarray}
\frac{d u_n}{dt}= i (a k_n u_{n+2}u_{n+1}^*+b
k_{n-1}u_{n+1}u_{n-1}^* \nonumber \\ -c k_{n-2}u_{n-1}u_{n-2}) -
\nu k_n^2 u_n +f \delta_{n,0}\ . \label{sabra}
\end{eqnarray}
where $f \delta_{n,0}$ is the forcing acting only on the first shell,
$\nu$ is the kinematic viscosity and $u_n^*$ is the complex conjugate of
$u_n$. We use $a=1$, $b=-0.5$, and $c=-0.5$. 
With this choice of the
parameters, the model satisfies both energy and helicity conservation 
in the inviscid limit.

Multiplying Eq.~(\ref{sabra}) by $u_n^*$, the complex conjugate of $u_n$, 
and taking the average of the resulting equation and its
complex conjugate, we obtain the energy budget for the $n$th shell:
\begin{equation}
\label{energy} \frac{1}{2} \frac{d |u_n|^2}{dt} = F_u(k_n)-F_u(k_{n+1})
-\nu k_n^2 |u_n|^2 + \Re(f u_0^*) \delta_{n,0} \ ,
\end{equation}
where
\begin{equation}
F_u(k_n) \equiv k_{n} \Im \big(u_{n-1}^*u_n^* u_{n+1} + \frac{1}{4} u_n u_{n-1}^*u_{n-2}^*
 \big)\ 
\label{Fn}
\end{equation}
Here $\Re$ and $\Im$ represent the real and imaginary parts, respectively.
The physical meaning of the different terms in the right hand side of
Eq.~(\ref{energy}) is clear. The third term 
is the rate of energy dissipation in the $n$th shell due to viscosity,
the last term is the power input due to external forcing,
and $F_n$ is the rate of 
energy transfer from the $(n-1)$th shell to the $n$th shell. 
In stationary state and in the inertial range at which the external forcing
is not acting and energy dissipation is negligible,
Eq.~(\ref{energy}) gives
\begin{equation}
0 \approx \langle F_u(k_n) \rangle - \langle F_u(k_{n+1}) \rangle
\end{equation}
which implies that 
$\langle F_u(k_n) \rangle$ is independent of $k_n$ in the inertial range, 
which is the statement of energy cascade.
The ensemble averages are evaluated as time averages when the system is 
in the stationary state.
In the shell model, the local energy transfer rate $\Pi_r$ 
can thus be naturally identified as $F_u(k_n)$ with $k_n=1/r$: 
\begin{equation}
\Pi_r \to F_u(k_n)
\label{shellPir}
\end{equation}

As for the local energy dissipation rate, it has to be defined accordingly in the shell 
model. From Eq.~(\ref{energy}), the total energy dissipation rate (at time $t$) 
in the shell model is 
\begin{equation}
\epsilon(t) = \sum_{n=0}^{N-1} \nu k_n^2 |u_n|^2
\label{epsilonshell}
\end{equation} 
We define the analog of the local energy dissipation rate 
$\epsilon_n$ in the shell model as
\begin{equation}
\epsilon_r \to \epsilon_n \equiv \sum_{n'=n}^{N-1} 
\nu k_{n'}^2 |u_{n'}|^2  \ ,
\label{shellepsilonr}
\end{equation}
again with $k_n=1/r$.
With this definition, we have the nice result of $\epsilon_n(t)$ approaching $\epsilon(t)$
in the limit of $n \to 0$ or equivalently in the limit of $r \to L$.
Summation of Eq.~(\ref{energy}) from $n>1$ to $N-1$ gives
\begin{equation}
\label{energy_sum}\frac{1}{2} \frac{d}{dt} \sum_{n'=n}^{N-1} |u_n'|^2 =
F_u(k_n) - \epsilon_n \ ,
\end{equation}
Thus in the stationary state, we have 
\begin{equation}
\langle F_u(k_n) \rangle = \langle \epsilon_n \rangle 
\end{equation}
but $F_u(k_n) \ne \epsilon_n$ and, in general, 
the two quantities can have different statistical features.

The velocity structure functions $Q_p(k_n)$ are statistical moments of $|u_n|$ and
$Q_p(k_n)$ scales with $k_n$ with scaling exponents $\gamma_p$:
\begin{equation}
Q_p(k_n) \equiv \langle |u_n|^p \rangle \sim k_n^{-\gamma_p}
\label{zetap}
\end{equation}
As reported in Ref.~\cite{Sabra}, the values of $\gamma_p$ are close to those obtained
in experiments and deviate from the K41 values of $p/3$.
In this work, we are interested in examining the validity of the RSHs
for accounting this anomalous scaling behavior of the velocity structure functions.
With $r = 1/k_n$, the shell-model analog of Kolmogorov's
RSH~[Eq.~(\ref{KolRSH})] is written as:
\begin{equation}
u_n = \phi \ \epsilon_n^{1/3} k_n^{-1/3} \label{KolRSHshell} 
\end{equation}
where $\phi$ is a random variable independent of $n$ and statistically independent of $\epsilon_n$.
Equation~(\ref{KolRSHshell})
implies that the conditional velocity structure functions at
fixed values of $\epsilon_n$ would have simple K41 scaling:
\begin{equation}
{\tilde Q}_p(k_n)
\equiv \langle |u_n|^p  \ \big| \epsilon_n \rangle \sim k_n^{-p/3} \label{Kol}
\end{equation}
For simplicity of the notation, we suppress the dependence on $\epsilon_n$ in
the conditional velocity structure functions $\tilde Q_p$. The notations for
the other conditional structure functions will be simplified in the same fashion.
Similarly, the shell-model analog of Kraichnan's RSH reads:
\begin{equation}
u_n = \psi |F_u(k_n)|^{1/3} k_n^{-1/3} 
\label{KraRSHshell} 
\end{equation}
where $\psi$
is a random variable independent of $n$ and statistically independent of $F_u(k_n)$.
One implication of Eq.~(\ref{KraRSHshell}) is that 
$\langle |u_n|^{3p} \rangle$ and $\langle |F_u(k_n)|^p \rangle k_n^{-p}$ 
have the same scaling behavior. 
This implication has indeed been confirmed in the 
GOY~\cite{KLWB95} and Sabra~\cite{Sabra} models.
From the definition of $F_u$ [see Eq.~(\ref{Fn})],
it is not surprising that $|F_u(k_n)|$ would have the same scaling behavior 
as $k_n |u_n|^3$, and thus the above implication could well be a direct
consequence of the definition of $F_u(k_n)$ in the shell model.
However, we emphasize that Eq.~(\ref{KraRSHshell}) is not merely a statement of the
relation of the scaling 
exponents of $\langle |u_n|^{3p} \rangle$ and $\langle |F_u(k_n)|^p \rangle$. 
In particular, it has another important implication, namely,
the conditional velocity structure functions 
at fixed values of $F_u(k_n)$ would have simple K41 scaling: 
\begin{equation}
Q_p^*(k_n) 
\equiv \langle |u_n|^p  \ \big| F_u(k_n) \rangle \sim k_n^{-p/3} \label{Kra} 
\end{equation}
Equation~(\ref{Kra}) is nontrivial 
if $u_n$ fluctuates even when $F_u(k_n)$ is fixed at some given value,
which would be the case 
when $\psi$ is a random variable with a certain probability distribution rather than a number of
fixed value.

\section{Shell model for homogeneous turbulent convection}
\label{Brandenburg}

Homogeneous turbulent convection has been proposed~\cite{Orszag}
as a convective flow in a box, with periodic
boundary conditions, driven by a constant temperature gradient along the
vertical direction. In Boussinesq approximation,
the equations of motion read~\cite{Ultimate}:
\begin{eqnarray}
\frac{\partial {\vec u}}{\partial t}
+ {\vec u} \cdot {\vec \nabla} {\vec u}
&=& -{\vec \nabla} p + \nu \nabla^2 {\vec u} + \alpha g \theta {\hat z}
\label{convection}  \\
\frac{\partial \theta}{\partial t}
+ {\vec u} \cdot {\vec \nabla} \theta &=& \kappa \nabla^2 \theta + \beta u_z \ \ 
\label{Teqn}
\end{eqnarray}
with ${\vec \nabla} \cdot {\vec u} = 0$.
Here, $p$ is
the pressure divided by the density,
$\theta = T- (T_0 - \beta z)$ is the deviation of temperature from
a linear gradient $-\beta$,
$T_0$ is the mean temperature of the fluid, and
${\hat z}$ is a unit vector in the vertical direction.

A shell model for homogeneous turbulent convection
driven by a temperature gradient was proposed by 
Brandenburg~\cite{Brandenburg}.
In Brandenburg's shell model,
the velocity and temperature variables are real.
We denote the velocity variable as $v_n$, to distinguish it from $u_n$ in the Sabra model,
and the temperature variable as $\theta_n$. The equations of motions are:
\begin{eqnarray}
\nonumber
\frac{d v_n}{d t} + \nu k_n^2 v_n
&=& \alpha g \theta_n + Ak_n(v_{n-1}^2-\lambda v_nv_{n+1}) \\
 &+& Bk_n(v_nv_{n-1}- \lambda v^2_{n+1}) 
\label{un}  \\
\nonumber
\frac{d \theta_n}{d t} + \kappa k_n^2 \theta_n
&=& \beta v_n + {\tilde A} (k_n(v_{n-1}\theta_{n-1}-k_{n+1} v_n\theta_{n+1}) \\
&+& {\tilde B}( k_n v_n \theta_{n-1}-k_{n+1} v_{n+1}\theta_{n+1})  \ \
\label{thetan}
\end{eqnarray}
where $A$, $B$, $\tilde A$, and $\tilde B$ 
are positive parameters.
Earlier work showed that the scaling behavior
depends only on the ratio $B/A$~\cite{Brandenburg}.
It has been recently shown~\cite{ChingCheng} that buoyancy drives the dynamics 
and affects the statistics of 
$v_n$ and $\theta_n$ when $B/A$ is greater than some critical value of about 2.
When buoyancy is driving the dynamics, 
energy is transferred from small to large scales on
average~\cite{Brandenburg}. As a result, a linear damping term 
$-f_0 v_n \delta_{n,0}$~\cite{ST95,CMMV2002} 
has to be added to Eq.~(\ref{un}) for the system to achieve stationarity.

In Bousinessq approximation, entropy is proportional to the volume
integral of the temperature variance. Entropy in the $n$th shell is, therefore,
defined as ${\cal S}_n \equiv \theta_n^2/2$. By studying the entropy
budget obtained from
Eq.~(\ref{thetan}) upon multiplication by $\theta_n$:
\begin{equation}
\frac{d {\cal S}_n}{dt} = F_\theta(k_n)-F_\theta(k_{n+1})-\kappa k_n^2 \theta_n^2 +
\beta v_n \theta_n
\label{Sn}
\end{equation}
we get the rate of entropy transfer or entropy flux from $(n-1)$th to $n$th shell as:
\begin{equation}
F_\theta(k_n) \equiv k_n({\tilde A}v_{n-1}+
{\tilde B}v_n)\theta_{n-1}\theta_n
\end{equation}
In the stationary state and in the intermediate range in which $\langle v_n \theta_n \rangle$ and
the entropy dissipation $\kappa k_n^2 \langle \theta_n^2 \rangle $ are negligible,
Eq.~(\ref{Sn}) gives:
\begin{equation}
0 \approx \langle F_\theta(k_n) \rangle -
\langle F_\theta(k_{n+1}) \rangle
\end{equation}
Thus $\langle F_\theta(k_n) \rangle$ is independent of $k_n$ in the intermediate range of scales,
which is the statement of entropy cascade.
We naturally identify the local entropy transfer rate $\Pi^{\theta}_r$ as
$F_\theta(k_n)$ with $k_n = 1/r$:
\begin{equation}
\Pi_r^{\theta} \to F_\theta(k_n)
\label{shellPithetar}
\end{equation}
In the Brandenburg shell model, the total entropy dissipation rate (at time $t$) is:
\begin{equation}
\chi(t) = \sum_{n=0}^{N-1} \kappa k_n^2 \theta_n^2
\label{shellchi}
\end{equation}
Thus as in Eq.~(\ref{shellepsilonr}), we define  
the analog of the local entropy dissipation rate in the shell model, $\chi_n$, as
\begin{equation}
\chi_r \to \chi_n \equiv \sum_{n'=n}^{N-1} \kappa k_{n'}^2 \theta_{n'}^2
\label{shellchir}
\end{equation}
again with $k_n=1/r$.

The velocity and temperature structure functions, defined by
\begin{eqnarray}
S_p(k_n) &\equiv& \langle |v_n|^p \rangle \sim k_n^{-\zeta_p} \\
R_p(k_n) &\equiv& \langle |\theta_n|^p \rangle \sim k_n^{-\xi_p} \ ,
\end{eqnarray}
have been studied recently~\cite{ChingCheng} and found to have anomalous scaling behavior.
We would like to examine whether this anomalous scaling can be
understood using the RSHs extended to turbulent convection.
The shell-model analogs of the RSH of Kolmogorov extended to turbulent convection, 
Eqs.~(\ref{KolRSHActiveU}) and (\ref{KolRSHActiveT}), are:
\begin{eqnarray}
v_n &=& \Phi_v (\alpha g)^{2/5} \chi_n^{1/5} k_n^{-3/5} 
\label{shellKolU}\\
\theta_n &=& \Phi_\theta (\alpha g)^{-1/5} \chi_n^{2/5} k_n^{-1/5} 
\label{shellKolT}
\end{eqnarray}
which imply that
\begin{eqnarray}
{\tilde S}_p(k_n) &\equiv& \langle |v_n|^p \ \big| \chi_n \rangle \sim k_n^{-3p/5} 
\label{shellcondKolU}\\
{\tilde R}_p(k_n) &\equiv& \langle |\theta_n|^p \ \big| \chi_n \rangle \sim k_n^{-p/5} 
\label{shellcondKolT}
\end{eqnarray}
Similarly the shell model analogs of the RSH of Kraichnan extended to turbulent convection,
Eqs.~(\ref{KraRSHActiveU}) and (\ref{KraRSHActiveT}), are~\cite{ChingCheng}:
\begin{eqnarray}
v_n &=& \Psi_v (\alpha g)^{2/5} |F_\theta(k_n)|^{1/5} k_n^{-3/5} 
\label{shellKraU}\\
\theta_n &=& \Psi_\theta (\alpha g)^{-1/5} |F_\theta(k_n)|^{2/5} k_n^{-1/5} 
\label{shellKraT}
\end{eqnarray}
and they imply 
\begin{eqnarray}
S^*_p(k_n) &\equiv& \langle |v_n|^p \ \big| F_\theta(k_n) \rangle \sim k_n^{-3p/5} 
\label{shellcondKraU}\\
R^*_p(k_n) &\equiv& \langle |\theta_n|^p \ \big| F_\theta(k_n) \rangle \sim k_n^{-p/5}
\label{shellcondKraT}
\end{eqnarray}
Equations~(\ref{shellcondKraU}) and (\ref{shellcondKraT}) have been confirmed 
in Ref.~\cite{ChingCheng} thus
supporting the validity of the extension of Kraichnan's RSH to turbulent
convection~[Eqs.~(\ref{shellKraU}) and (\ref{shellKraT})] in the
Brandenburg model. The validity of the extension of Kolmogorov's RSH to turbulent 
convection~[Eqs.~(\ref{shellKolU}) and (\ref{shellKolT})] in the Brandenburg model 
will be examined and discussed in Sec.~\ref{result}.

\section{Results and discussions}
\label{result}
\subsection{Validity of the RSHs of Kolmogorov and Kraichnan in Sabra model}
\label{result1}

We numerically integrate Eq.~(\ref{sabra}) using the
fourth order Runge-Kutta method. Starting with
$u_n = x k_n^{-1/3}$, where $x$ is a random variable,
we evolve the equations in time for a short period of time.
The results so obtained, with the phases randomized, are used as the 
initial data for the actual runs. Following Ref.~\cite{Sabra}, we 
take the external forcing $f$ as a time correlated noise with exponential 
correlation and correlation time $\tau$  such that it follws the
evolution equation:
\begin{eqnarray}
\nonumber
&& f(t+dt) \\
&=& f(t)E + \sigma\sqrt{-2(1-E^2) \log_{10}(\rho_1)} \exp(i2\pi\rho_2) \ \  \qquad
\label{df} 
\end{eqnarray}
where $E=\exp(-dt/\tau)$, $\sigma$ is the standard deviation of $f$, and $\rho_1$
and $\rho_2$ are two uniform random numbers between 0 and 1.
We use $\sigma = 0.01$, $\tau=1$, and $f(t=0)=5(1+i) \times 10^{-3}$.

\vspace{0.8cm}

\begin{figure}[bth]
\centering \epsfig{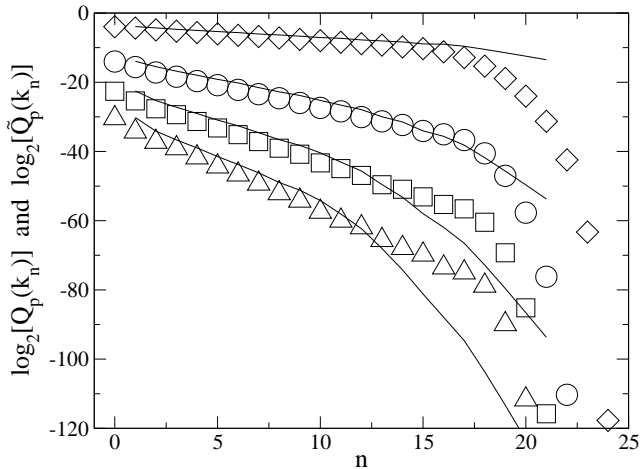}
\vspace{0.2cm}
\caption{The velocity structure functions $Q_p(k_n)$
as a function of the shell index $n$ with $p=1$~(diamonds), $p=4$~(circles),
$p=7$~(squares), and $p=10$~(triangles). The solid lines are the conditional
velocity structure functions $\tilde Q_p(k_n)$ at given values of 
$\epsilon_n$ with $10^{-4}\le \epsilon_n\le 3 \times 10^{-4}$ 
for the same values of $p$.}
\label{fig1}
\end{figure}

\vspace{0.5cm}

\begin{figure}[bth]
\centering \epsfig{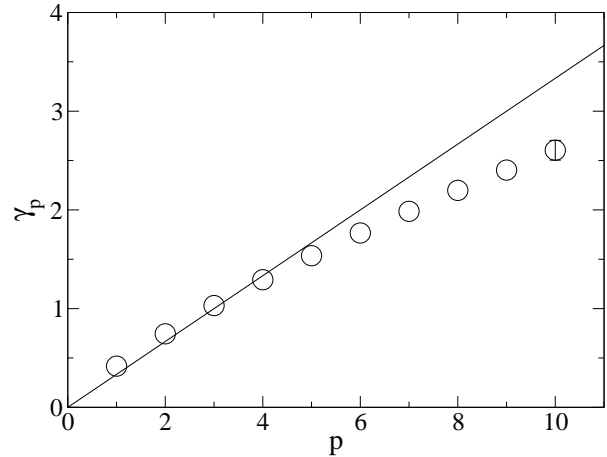}
\caption{The scaling exponents $\gamma_p$~(circles)
for $Q_p(k_n)$ as a function of $p$. The solid line is the K41 result of $p/3$. 
The error increases with $p$ and we show the largest error for $p=10$.}
\label{fig2}
\end{figure}

We calculate the velocity structure functions $Q_p$, and the conditional velocity
structure functions $\tilde Q_p$ and $Q^*_p$ respectively at fixed values of $\epsilon_n$
and $F_u(k_n)$. The statistics are collected by averaging over a time of approximately 
$1000$ eddy turnover time of the largest scales.
The scaling behavior of the model does not depend on the value of $k_0$ and
is also independent of the value of $\nu$ when $\nu$ is small enough.
We have used several different values of $N$.
For the results reported below, we have used $k_0=2$, $\nu=2 \times 10^{-8}$ and $N=24$,
and $\nu=3.2 \times 10^{-7}$ and $N=21$.
Some of the results for the velocity structure functions $Q_p$ are shown in Fig.~\ref{fig1}.
We observe that they exhibit nice power-law dependence on $k_n$ for $3<n<16$.
The scaling exponents $\gamma_p$ obtained
are plotted in Fig.~\ref{fig2}.
These values, which are in good agreement with the values reported in the literature~\cite{Sabra},
deviate from $p/3$ demonstrating clearly that the velocity structure functions
exhibit anomalous scaling.

\vspace{1cm}

Our aim is to examine the validity of Kolmogorov and Kraichnan's RSHs
[Eqs.~(\ref{KolRSHshell}) and (\ref{KraRSHshell})] in accounting for the origin of the anomalous
scaling behavior by examining the validity of
Eqs.~(\ref{Kol}) and (\ref{Kra}).
To do so, we study the scaling exponents of 
$\tilde Q_p(k_n)$ and $Q^*_p(k_n)$, defined by:
\begin{eqnarray}
\tilde Q_p(k_n) &\sim& k_n^{-\tilde \gamma_p} \label{zetastar}  \\
Q^*_p (k_n) &\sim& k_n^{-{\gamma}^*_p} \ ,
\label{zetatilde}
\end{eqnarray}
and check directly whether
$\tilde \gamma_p$ and $\gamma^*_p$ agree with $p/3$.

\begin{figure}[t]
\centering \epsfig{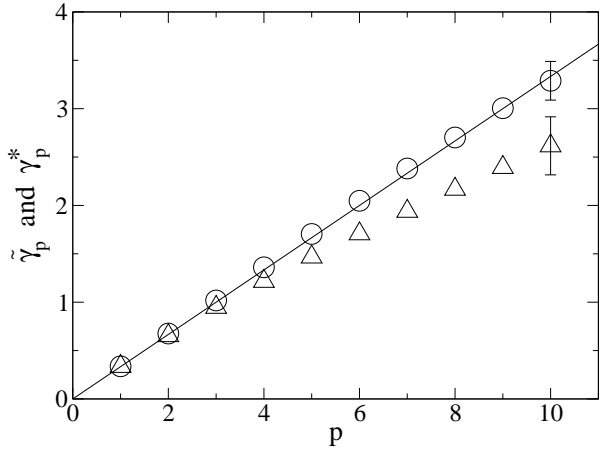}
\caption{The scaling exponents $\tilde \gamma_p$~(triangles) an $\gamma^*_p$~(circles)
of $\tilde Q_p(k_n)$ and $Q^*_p(k_n)$ respectively as a function of $p$. 
The solid line is the K41 result of $p/3$ and the largest errors for $p=10$ are shown.}
\label{fig3}
\end{figure}

To calculate the conditional velocity structure functions $\tilde Q_p$ at 
fixed values of $\epsilon_n$, we average only those $|u_n|^p$ when
the value of $\epsilon_n$ is within a given narrow range of values.
The results for $\tilde Q_p$ are also shown in Fig.~\ref{fig1}.
It can be seen that $\tilde Q_p$ are different from $Q_p$.
Moreover, the scaling range of $\tilde Q_p$ is shorter than that of
$Q_p$. Thus $|u_n|$ is correlated with $\epsilon_n$. Yet the  
the scaling exponents $\tilde \gamma_p$ and $\gamma_p$ are close to 
one another. In particular, as shown in Fig.~\ref{fig3}, 
$\tilde \gamma_p$ continue to deviate from $p/3$,
demonstrating that Eq.~(\ref{Kol}) and thus Eq.~(\ref{KolRSHshell}) is invalid.  

\vspace{0.8cm}

\vspace{0.2cm}

\begin{figure}[bth]
\centering \epsfig{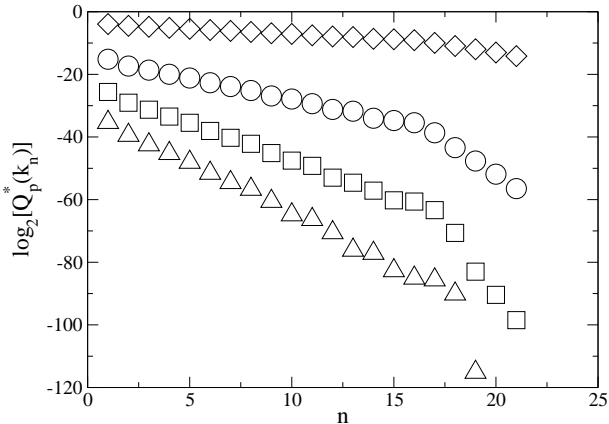}
\vspace{0.2cm}
\caption{The conditional velocity structure functions $Q^*_p(k_n)$ at fixed values
of $F_u(k_n)$ with $10^{-4}\le F_u(k_n)\le 3 \times 10^{-4}$.
Same symbols as in Fig.~\ref{fig1}.}
\label{fig5}
\end{figure}

\begin{figure}[bth]
\centering \epsfig{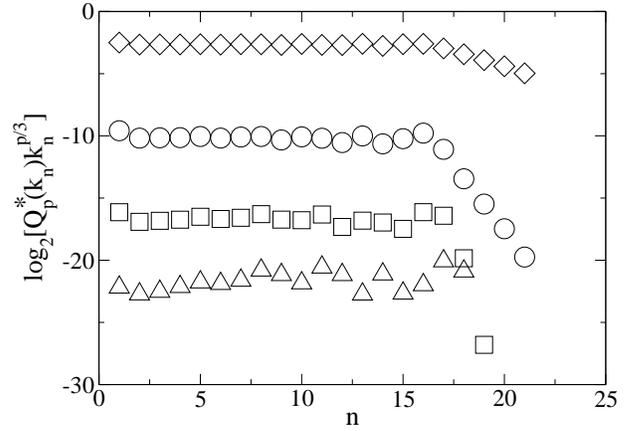}
\vspace{0.2cm}
\caption{Compensated plots of $Q^*_p(k_n) k_n^{p/3}$ versus $n$.
Same symbols as in Fig.~\ref{fig5}.}
\label{fig7}
\end{figure}

Similarly, to calculate the conditional velocity structure functions $Q^*_p$
at fixed values of $F_u(k_n)$, we average only those $|u_n|^p$ when
$F_u(k_n)$ assumes values in the same given narrow range.
Some of the results are shown in Fig.~\ref{fig5}. It can be seen that $Q^*_p$
are again different from $Q_p$ thus showing that $u_n$ and $F_u(k_n)$ are correlated.
Moreover, the scaling range of $Q^*_p$ is comparable to that of $Q_p$.
In Fig.~\ref{fig3}, we see that the scaling exponents $\gamma^*_p$
are consistent with $p/3$. We also
present the compensated plots of $Q^*_p(k_n) k_n^{p/3}$ in Fig.~\ref{fig7}.
It can be seen that the data points are consistent with being independent of $n$ in the
scaling range, thus demonstrating directly the scaling behavior of
$Q^*_p \sim k_n^{-p/3}$. These results thus confirm Eq.~(\ref{Kra}).

\vspace{0.5cm}

\vspace{0.5cm}

\begin{figure}[bth]
\centering \epsfig{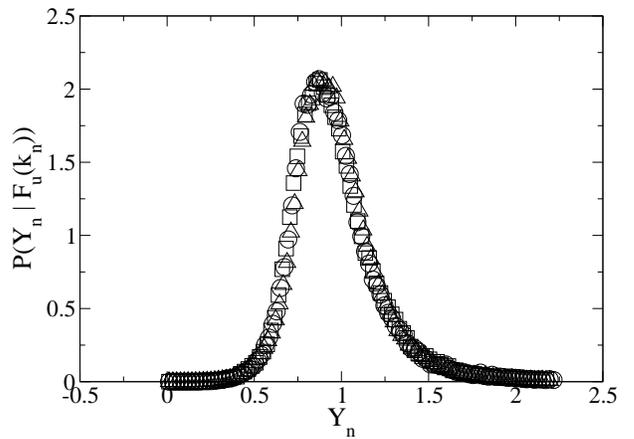}
\vspace{0.2cm}
\caption{The conditional probability density functions $P(Y_n \big| F_u(k_n))$
for $n~=~13$~(circles), $n~=~15$~(squares), $n~=~17$~(triangles).}
\label{fig8}
\end{figure}

As discussed in Sec.~\ref{Sabra}, it is important to check that
$u_n$ indeed fluctuates even when $F_u(k_n)$ is fixed at some given values.
Thus we study the conditional probability density functions of $u_n$ at
fixed values of $F_u(k_n)$ and confirm that the distributions indeed have 
a finite extent~[see Fig.~\ref{fig8}]. Moreover, the simple scaling behavior of
$Q^*_p$ implies that the conditional probability density functions 
$P(Y_n \big| F_u(k_n))$, with
$Y_n \equiv |u_n|/{\sqrt{\langle u_n^2 \big | F_u(k_n) \rangle}}$,
 are scale invariant or $n$-independent.
This is indeed the case as seen in Fig.~\ref{fig8}.
Our results thus show that Kraichnan's RSH
[Eq.~(\ref{KraRSHshell})] is valid in the Sabra shell model.

\vspace{1cm}

\vspace{1cm}
\subsection{Validity of extensions of Kolmogorov and Kraichnan's RSHs to turbulent convection 
in Brandenburg model}
\label{result2}

\begin{figure}[bth]
\centering \epsfig{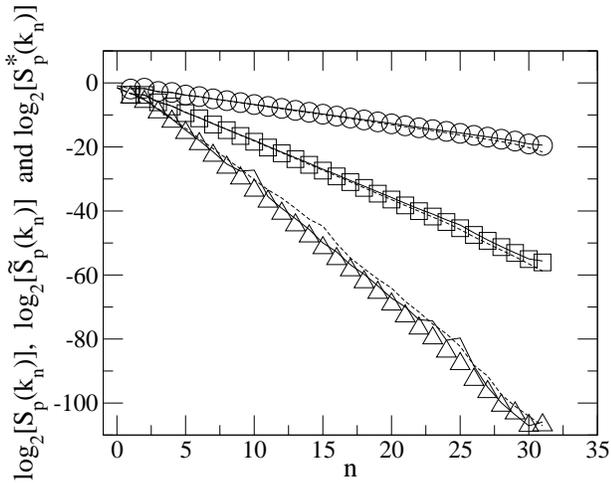}
\vspace{0.2cm}
\caption{$S_p$ for $p=1$~(circles), $p=3$~(squares), $p=6$~(triangles). The solid 
and dashed lines are respectively $\tilde S_p$ and 
$S^*_p$ for the same values of $p$.}
\label{fig9}
\end{figure}

\begin{figure}[bth]
\centering \epsfig{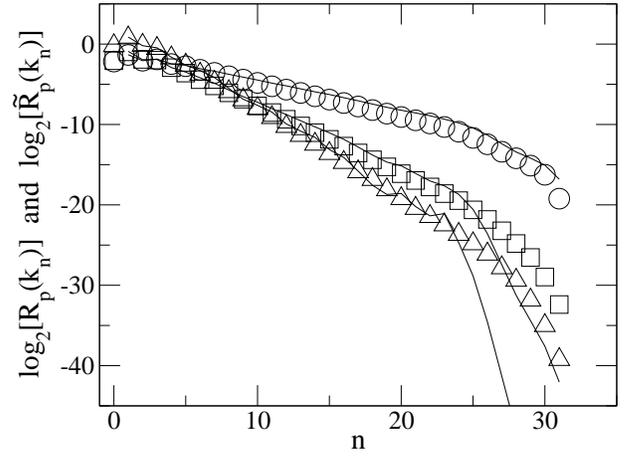}
\vspace{0.2cm}
\caption{$R_p$ for $p=2$~(circles), $p=5$~(squares), $p=8$~(triangles). The solid lines
are $\tilde R_p$ for the same values of $p$.}
\label{fig10}
\end{figure}

We numerically integrate Eqs.~(\ref{un}) and (\ref{thetan}) using
fourth-order Runge-Kutta method with an initial condition of $v_n=\theta_n=0$ except for
a small perturbation of $\theta_n$ in an intermediate value of $n$. We use $k_0=1$,
$A=0.01, B=1$, $\beta = 1$, $\tilde A=\tilde B=1$, $\alpha g=1$,  $\nu = 5 \times 10^{-17}$,
$\kappa = 5 \times 10^{-15}$, $f_0=0.5$, and $N=32$. 
We calculate the velocity and temperature structure functions $S_p$ and $R_p$
when the system is in stationary state. 
The results for $S_p$ and $R_p$ are shown in Figs.~\ref{fig9} and \ref{fig10}
respectively. It can be seen that they all have good scaling behavior.

\vspace{0.5cm}

\vspace{1.0cm}

\begin{figure}[bth]
\centering \epsfig{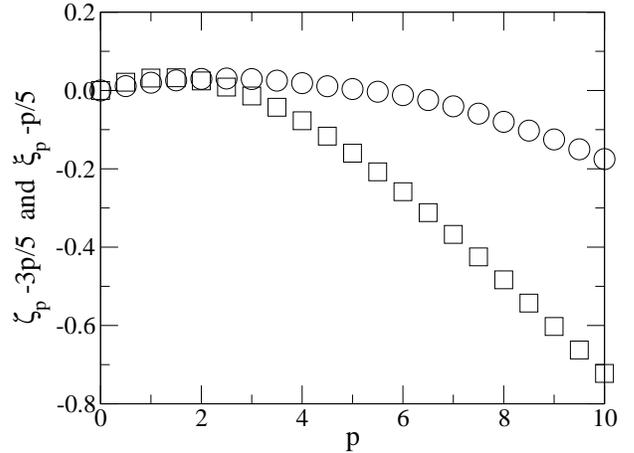}
\vspace{0.2cm}
\caption{The deviations of $\zeta_p$ and $\xi_p$ from the BO values: 
$\zeta_p-3p/5$~(circles) and $\xi_p -p/5$~(squares).}
\label{fig12}
\end{figure}

\begin{figure}[bth]
\centering \epsfig{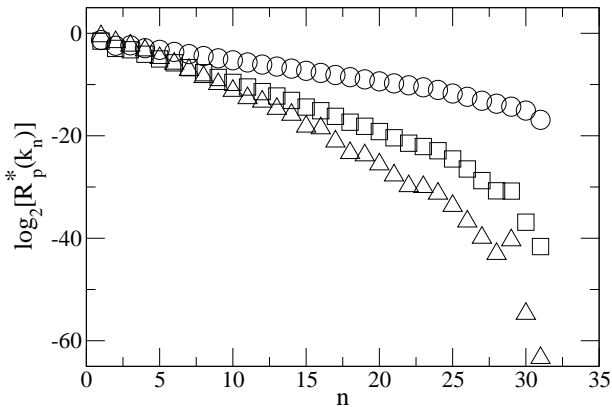}
\vspace{0.2cm}
\caption{The conditional temperature structure functions $R_p^*$ at fixed values
of $F_\theta(k_n)$ for $p=2$~(circles), $p=5$~(squares), $p=8$~(triangles).}
\label{fig11}
\end{figure}

In an earlier study~\cite{ChingCheng}, we have studied the scaling
behavior of an active scalar using the Brandenburg model and found
that $\zeta_p$ and $\xi_p$ are
given by the BO values plus corrections~(see Fig.\ref{fig12}). This
shows that $S_p$ and $R_p$ have anomalous scaling behavior.
In the same study, we have checked directly that the scaling exponents
of $S^*_p$ and $R^*_p$, denoted by $\zeta^*_p$ and $\xi^*_p$,
have the BO values of $3p/5$ and $p/5$, respectively, thus
confirming the validity of 
Eqs.~(\ref{shellcondKraU}) and (\ref{shellcondKraT}).
To calculate $S_p^*$ and $R_p^*$,
we average $|v_n|^p$ and $|\theta_n|^p$ only when $F_\theta(k_n) = 0.2 \pm 10^{-4}$.
We show $S_p^*$ also in Fig.~\ref{fig9}. It can be seen that $S_p^*$ are close to $S_p$ but with
a slightly steeper decrease with $k_n$. Results for $R_p^*$ are shown in Fig.~\ref{fig11} in which
the steeper decrease with $k_n$ can be seen more clearly. 
The validity of Eqs.~(\ref{shellcondKraU}) and (\ref{shellcondKraT}) indicates
the validity of the extensions of Kraichnan's RSH to turbulent
convection in the Brandenburg model. 

\vspace{1cm}

\vspace{0.5cm}

\begin{figure}[bth]
\vspace{0.5cm}
\centering \epsfig{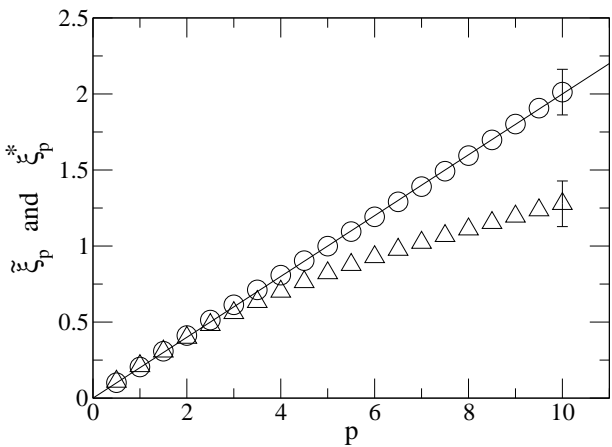}
\vspace{0.2cm}
\caption{Comparison of $\tilde \xi_p$~(triangles) and $\xi_p^*$~(circles) 
with the BO values of $p/5$~(solid line). The largest errors for $p=10$ are shown.}
\label{fig13}
\end{figure}

With these results,  we expect that
the extensions of Kolmogorov's RSH to turbulent convection
would be invalid in the Brandenburg model. In particular,
we expect Eqs.~(\ref{shellcondKolU}) and (\ref{shellcondKolT}) to be invalid. To show this, we
calculate $\tilde S_p$ and $\tilde R_p$ by averaging $|v_n|^p$ and $|\theta_n|^p$
only when $\chi_n = 0.2 \pm 10^{-4}$. Results for $\tilde S_p$ and $\tilde R_p$
are shown also in Figs.~\ref{fig9} and \ref{fig10}, respectively.
We find that both $\tilde S_p$ and $\tilde R_p$ are very close to $S_p$ and $R_p$
with $\tilde R_p$ having a shorter scaling range than $R_p$. The scaling exponents
of $\tilde S_p$, denoted by $\tilde \zeta_p$, are close to $\zeta_p$ which are close
to $3p/5$ within errors.
So we focus on the scaling exponents $\tilde \xi_p$ of $\tilde R_p$. In Fig.~\ref{fig13},
we compare $\tilde \xi_p$ with $\xi_p^*$ and the BO values of $p/5$
and see clearly that, as expected, Eq.~(\ref{shellcondKolT}) does not hold.

\vspace{0.5cm}

\section{Conclusions}
\label{conclusion}

One longstanding goal in turbulence research is to 
understand, from first principles, the origin of the anomalous scaling
of the fluctuating physical quantities.
An important idea was proposed by Kolmogorov in his refined theory~\cite{RSH},
which attributes the origin of the anomalous scaling behavior of the velocity fluctuations in
homogeneous and isotropic turbulence to variations of the local energy dissipation rate.
It was pointed out later by Kraichnan~\cite{Kraichnan} 
that it is more appropriate to replace the local
energy dissipation rate by the local energy transfer rate. We refer these two ideas to
as Kolmogorov's and Kraichnan's RSHs. Both of them can be extended to account for
the anomalous scaling behavior of the velocity and temperature fluctuations in 
turbulent convection in which temperature acts as an active scalar. 
Specifically, in the extension of Kolmogorov's RSH to turbulent convection,
the local entropy (or temperature variance) dissipation rate plays the role of
the local energy dissipation rate. Similarly, in the extension of Kraichnan's RSH to turbulent
convection, the local entropy transfer rate plays the part of the local energy transfer rate.

In this paper, we have examined the validity of Kolmogorov's and Kraichnan's RSHs
and their extensions to turbulent convection respectively in the Sabra shell model of
homogeneous and isotropic turbulence~\cite{Sabra} and in the Brandenburg shell model of 
homogeneous turbulent convection~\cite{Brandenburg}. The validity of Kolmogorov's RSH
has been examined in previous studies. These studies mainly
focussed on the statistical correlation of the velocity difference
and the local energy dissipation rate and the relation between the scaling exponents
of the velocity structure functions and the moments of the local 
energy dissipation. The validity of Kraichnan's RSH is much less studied and again
the focus is on the relation between the scaling exponents of the velocity structure functions
and the moments of the local energy transfer rate.
An important consequence of Kolmogorov's or Kraichnan's RSH is that the
conditional velocity structure functions at given values of the
local energy dissipation rate or the local energy transfer rate
would have simple K41 scaling behavior. Similarly, an important
consequence of the extensions of Kolmogorov's or Kraichnan's RSH to turbulent
convection is that the conditional velocity and temperature structure functions
at given local entropy dissipation rate or the local entropy transfer rate
would have simple BO scaling behavior. Our approach is to check directly these consequences.
In the shell models, the local energy or entropy transfer rate is easily identified
with the shell-to-shell energy or entropy transfer rate. On the other hand,
the local energy or entropy dissipation rate has to be defined accordingly.
As shown in Fig.~\ref{fig3}, the scaling exponents $\gamma^*_p$ of 
the conditional velocity structure functions at given shell-to-shell energy transfer rate 
indeed have the K41 values of $p/3$ while the scaling exponents
$\tilde \gamma_p$ of the conditional velocity structure functions at 
given shell-model analog of the local energy dissipation rate 
continue to deviate from the K41 values. This result shows that Kraichnan's RSH but not 
Kolmogorov's RSH holds in the Sabra model. 
Similarly, as shown in Fig.~\ref{fig13}, we have found that the scaling 
exponents $\xi^*_p$ of the conditional temperature structure functions at 
given shell-to-shell entropy transfer rate
have the BO values of $p/5$ while the
scaling exponents $\tilde \xi_p$ of the conditional temperature structure functions at
given shell-model analog of the local entropy dissipation rate continue to deviate from 
the BO values. Our result thus shows that the extension of Kraichnan's RSH but not 
Kolmogorov's RSH to turbulent covection holds in the Brandenburg model. 
In summary, our work shows that Kraichnan's RSH and its extension to turbulent convection
hold in shell models of turbulence. It would be interesting to perform similar study 
in direct numerical simulations and in experimental investigations.

\newpage
\acknowledgments

We thank Roberto Benzi and Luca Biferale for stimulating discussions.
This work is supported in part by the Hong Kong Research Grants
Council (Grant No. CA05/06.SC01).

\end{document}